\documentclass{article}

\usepackage{PRIMEarxiv}
\usepackage{esvect}
\usepackage{subfig}
\usepackage[utf8]{inputenc} % allow utf-8 input
\usepackage[T1]{fontenc}    % use 8-bit T1 fonts
\usepackage{hyperref}       % hyperlinks
\usepackage{url}            % simple URL typesetting
\usepackage{booktabs}       % professional-quality tables
\usepackage{amsfonts}       % blackboard math symbols
\usepackage{nicefrac}       % compact symbols for 1/2, etc.
\usepackage{microtype}      % microtypography
\usepackage{lipsum}
\usepackage{fancyhdr}       % header
\usepackage{graphicx}       % graphics
\graphicspath{{media/}}     % organize your images and other figures under media/ folder

\title{Symmetry Breaking during Low-Temperature Domain Formation in Micron-sized Magnetite Crystals
}

\author{
  Yue Dong \\
  University of Oxford\\
  London Centre for Nanotechnology\\
  University College, London\\
  \texttt{yue.dong@lmh.ox.ac.uk} \\
   \And
  David Yang\\
  Brookhaven National Laboratory\\
  \texttt{dyang2@bnl.gov} \\
  \And
  Jialun Liu\\
   London Centre for Nanotechnology\\
   University College, London\\
  \texttt{jialun.liu.17@ucl.ac.uk} \\
    \And
  Aly Abdeldaim\\
Diamond Light Source\\
Harwell Science and Innovation Campus\\
  \texttt{aly.abdeldaim@diamond.ac.uk} \\
      \And
  Wei Wang\\
Brookhaven National Laboratory\\\
University of Science and Technology of China\\
  \texttt{wwpositron@ustc.edu.cn} \\
        \And
  Ian Robinson\\
  London Centre for Nanotechnology\\
  University College, London\\
Brookhaven National Laboratory\\\
  \texttt{i.robinson@ucl.ac.uk} \\
}

\begin{document}
\maketitle

\begin{abstract}
We report the results of synchrotron Bragg Coherent X-ray Diffraction Imaging (BCDI) experiments to investigate domain formation in a micron-sized magnetite crystal undergoing the Verwey transition at low temperature. A strong splitting of the measured 311 Bragg reflection was observed in the low-temperature phase, indicating the formation of domains. BCDI revealed  pronounced strain distributions, characterized by a clear layered stripe domain structure in real-space. Stripes were seen only along the [001] crystallographic direction, normal to the substrate surface direction, breaking the symmetry of the cubic high-temperature phase. It is argued that other domain directions were suppressed by the sample mounting orientation. More surprisingly, only a single domain orientation was observed, suggesting an additional symmetry-breaking influence originating from the shape of the crystal.
\end{abstract}

% keywords can be removed

\section{Introduction}
Magnetite (Fe$_{3}$O$_{4}$), a mixed-valence iron oxide\cite{Gleitzer1985}, is known for its correlated electronic and magnetic properties. It is a naturally occurring mineral, historically famous in ancient applications in compasses due to its strong magnetic moment which persists above room temperature. It has gathered attention for its electron transport (EET) mechanisms\cite{Liu2015}, phase transitions and charge ordering phenomena\cite{Lee2008}\cite{Senn2011}\cite{Lorenzo2008}, correlated electron-related biological systems\cite{Kirschvink1992}\cite{Kobayashi2024}, and is emerging as a key component in the development of metallic and magnetic nanocomposites\cite{Fatimah2021}.

The Verwey transition\cite{Kirschvink1992} \cite{Kobayashi2024} \cite{Fatimah2021} \cite{Verwey1939} \cite{Verwey1936} \cite{Joaquin2004} \cite{Friedrich2002} represents a sharp metal-insulator phase transition at the stoichiometry-dependent Verwey temperature (T$_{v}$ $\sim$ 125 K), intimately linked to a change in the material’s crystallographic structure and electron-lattice interactions. Above the Verwey transition temperature, magnetite adopts an inverse spinel cubic structure\cite{Verwey1936}\cite{Bragg1915}\cite{Lizumi1982} (space group Fd-3m), with Fe$^{3+}$ ions occupying the tetrahedral (A) sites and a random mixture of Fe$^{2+}$ and Fe$^{3+}$ ions located at the octahedral (B) sites. This structure results in metallic conductivity due to electron hopping between the Fe$^{2+}$ and Fe$^{3+}$ ions on the B-sites. 

As the temperature drops below the Verwey transition, magnetite undergoes a structural phase change from the cubic inverse spinel space group to a monoclinic or triclinic distorted phase\cite{Wright2011}\cite{Wright2002}\cite{Perversi2019}. This change is accompanied by charge ordering of the Fe$^{2+}$ and Fe$^{3+}$ ions to form “trimerons”, valence-ordered Fe$^{3+}$-Fe$^{2+}$-Fe$^{3+}$ linear structures\cite{Senn2011}, confining the electrons and leading to an insulating behavior at low temperatures. The transition involves both electronic localization and a significant lattice distortion, evidenced by a change in symmetry and unit cell dimensions. The higher electrical conductivity of the cubic high-temperature phase is explained by the random Fe$^{2+}$/Fe$^{3+}$ occupation, whereby mobile electrons can freely change the Fe site valency. The structure of the low-temperature phase, with trimeron bonds of charge ordering and strain, is adequately described as a monoclinic phase\cite{Wright2002}. When the crystal distorts into this structure it has the possibility of forming Martensitic domains\cite{Shuvalov1988}. Martensitic transformation is a diffusionless, solid-state phase transition that occurs in certain materials, particularly in metals and alloys like steel. It is characterized by a homogeneous movement, leading to a change in crystal structure without long-range diffusion. In this work, we will demonstrate the domain formation associated with T$_{v}$ and discuss how it could be influenced by geometric limits on the shape of the crystal through strain, contributing to the overall phase behavior. This represents one of the established mechanisms for modifying the physical properties of nanocrystals\cite{Meyers2006}.

While previous studies using X-ray diffraction\cite{Tabis2009}, electron microscopy\cite{Lindquist2019}, and neutron scattering\cite{Wright2000} have contributed valuable insights, there remain uncertainties involving how these transitions manifest at the nanoscale within individual micron-sized grains and the role of domain structures and lattice distortions associated with this phase transition. Electron microscopy has revealed significant domain and twin boundary formation at low temperatures upon crossing the Verwey transition\cite{Lindquist2019}, indicating that the transition is accompanied by charge ordering and complex strain-induced phenomena in magnetite crystals. Neutron powder diffraction data from a sample of polycrystalline Fe$_{3}$O$_{4}$ indicated a change in the thermal expansion and a change in crystallite extinction due to twinning at the transition\cite{Wright2000}.

These domain structures can further influence magnetite's electronic and magnetic properties, potentially stabilizing or suppressing certain phases at low temperatures. How strain accumulates within magnetite crystals as a signature of the monoclinic/triclinic phase and how it influences the electronic localization that defines the Verwey transition remains unclear. To address this here, we used Bragg Coherent X-ray Diffraction Imaging (BCDI)\cite{Robinson2009} to investigate micron-sized magnetite crystals undergoing the Verwey transition. BCDI is a technique used to image three-dimensional (3D) displacement fields within finite crystalline materials, with a spatial resolution of $\sim$10 nm. A coherent X-ray beam illuminates the sample, and a detector is placed in the far-field to capture the resulting diffraction pattern corresponding to a chosen crystallographic Bragg reflection. The sample is rotated about an axis, or “rocked,” over an angular range of $\sim$0.4$^\circ$ to move the Ewald sphere through the detector, capturing the 3D Bragg peak as a series of 2D diffraction patterns\cite{Robinson2009}\cite{Hofmann2017}\cite{Pfeifer2006} If the fringes in the resulting 3D diffraction pattern are oversampled, the pattern can be inverted to a 3D image of the sample using phase retrieval algorithms\cite{Robinson2009}\cite{Fienup1978}\cite{Chen2007}, or more recently, methods aided by deep neural network algorithms\cite{Yu2024}\cite{Wu2021}\cite{Masto2024}. Such images are complex, consisting of 3D spatial maps of amplitude and phase, $\Delta \phi(r)$, which correspond to the crystal's morphology and local displacement field, $\vv{u}(\vv{r})$ projected along the scattering vector, $\vv{Q}$, given by the equation: 

\begin{equation}
    \Delta \phi(\vv{r}) = \vv{u}(\vv{r})\cdot \vv{Q}
    \label{eq:1}
\end{equation}

BCDI is lensless and non-destructive, making it particularly suitable for investigating dynamic nanoscale phenomena in situ, such as annealing\cite{Yang2021}, dissolution\cite{Clark2015}, battery charging\cite{Liu2022}, catalysis\cite{Atlan2023} and phase transitions\cite{Diao2020}. Previous BCDI studies on magnetite have shown the evolution of the morphology and internal strain distribution of a $\sim$400 nm crystal undergoing oxidative dissolution in an acidic solution\cite{Yuan2019} and domain identification in manganite nanocrystal\cite{Mokhtar2024}. Here, we investigate the Verwey transition in a FIB-cut magnetite crystal by measuring 3D Bragg peaks as a function of temperature. A strong splitting of almost all these diffraction peaks was observed upon crossing the Verwey phase transition, indicating the formation of domains associated with the lower symmetry of the low-temperature phase\cite{Dhariwal2024}. After inverting the 311 Bragg peak, we observe a clear layered stripe domain structure and dramatic strain formation at temperatures below the Verwey transition temperature responsible for the splitting. The stripe domain structure corresponds to the streaking of the Bragg peak angled along the [001] crystallographic direction, normal to substrate surface direction. 

\section{Materials and Methods:}
\label{sec:headings}

\begin{figure}
    \centering
    \includegraphics[width=0.4\linewidth]{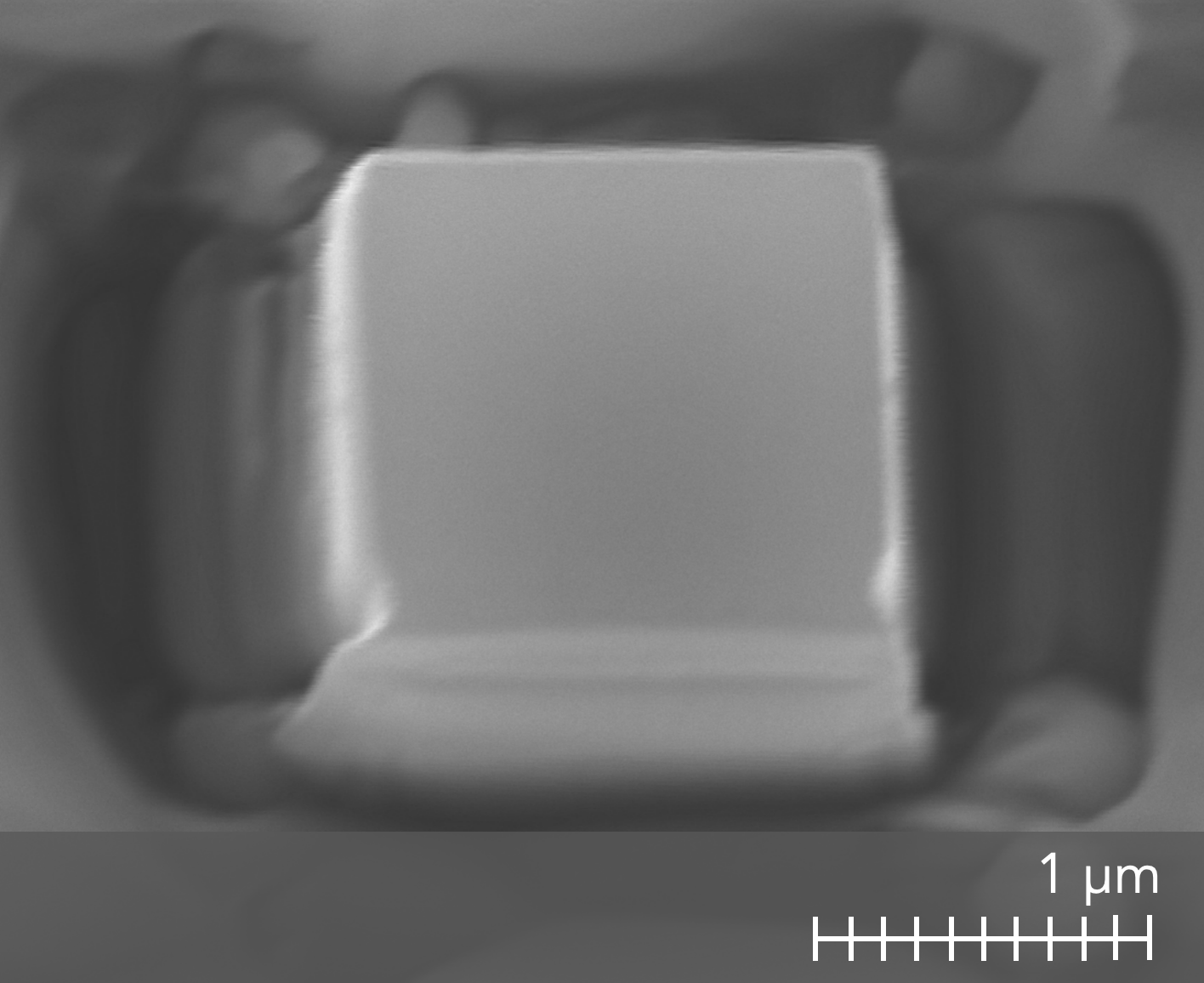}
    \caption{A top-down SEM image of the 1.6 $\mathrm{\mathrm{\mu} m}$ $\times$ 1.6 $\mathrm{\mu m}$ magnetite sample measured at 20 kV.}
    \label{fig:1}
\end{figure}

The magnetite Fe$_{3}$O$_{4}$ crystal was extracted from a bulk sample purchased from SurfaceNet GmbH using focused ion beam (FIB), a common sample fabrication technique where a gallium (Ga$^{+}$) primary ion beam sputters a small amount of material from a sample surface\cite{Hofmann2023}. A crystal fragment was lifted out from the parent crystal, trimmed to a cubic shape, and welded with Pt deposition onto a single crystal silicon substrate. Final polishing to the required dimensions for sufficient oversampling of the diffraction patterns was completed in situ on the substrate. The FIB magnetite crystal was imaged using scanning electron microscopy (SEM) at a voltage of 20 kV, shown in Figure $\ref{fig:1}$. This magnetite crystal was seen to be cube-shaped with a 1.6 $\mathrm{\mathrm{\mu} m}$ $\times$ 1.6 $\mathrm{\mu m}$ top surface. By orienting the crystal before the FIB preparation, its cube faces were aligned close to the cubic unit cell directions.

We performed an in situ BCDI temperature study of the FIB-cut Fe$_{3}$O$_{4}$ crystal at the I16 beamline at Diamond Light Source, Harwell Science and Innovation Campus, United Kingdom. The sample was mounted on the I16 displex cryostat in the vertical scattering geometry on the 6-circle kappa diffractometer, operating in a 4-circle mode\cite{Busing1967}. The diffraction pattern was indexed using the high-temperature crystal cubic unit cell with a$_{0}$ = 8.397$\mathring{A}$. The [004] direction, orthogonal to the silicon substrate plane, was found at $\chi$= 90$^\circ$ under symmetric diffraction conditions. The beamline’s 150 $\mathrm{\mathrm{\mu} m}$ $\times$ 50 $\mathrm{\mu m}$ focus reduced the effects of seeing fluctuations from the vibrations of the Displex cryostat, which we concluded were mostly translational. 

The experimental setup at I16 has an in-line Merlin (Medipix) detector at a distance $D$ = 1.3 m from the sample, with detector pixel size p = 55 $\mathrm{\mu m}$. The energy of the coherent X-ray beam was set at 9 keV (wavelength, $\lambda$ = 1.38$\mathring{A}$), allowing a maximum crystal size of $\lambda D/2p = 1.7 \mathrm{\mu m}$, given by the need for at least 2-fold oversampling\cite{Sayre1952}. The sample stage was covered by a thermal shield and a Beryllium window, allowing temperature scans from below 10 K to room temperature. The Verwey transition was found to occur at T $\sim$110 K for our sample, deviating from the predicted value of 125 K\cite{Verwey1939}. The Bragg peak was measured by collecting diffraction patterns through a series of 0.002$^\circ$ steps around the eta axis, using an exposure time of 1 s per step. The 113, 311, 511, 400, 004, and 404 reflections were measured at various temperatures. The most measurements were done on the 311 reflection (it had the strongest signal), around our sample’s Verwey transition temperature at 90 K, 100 K, 107 K, 108 K, 109 K, 110 K, 111 K, 115 K, and 120 K. Three consecutive scans were made at each temperature to ensure reproducibility. 

\section{Results}
\begin{figure}%
    \centering
    \subfloat[\centering]{{\includegraphics[width=10.3cm]{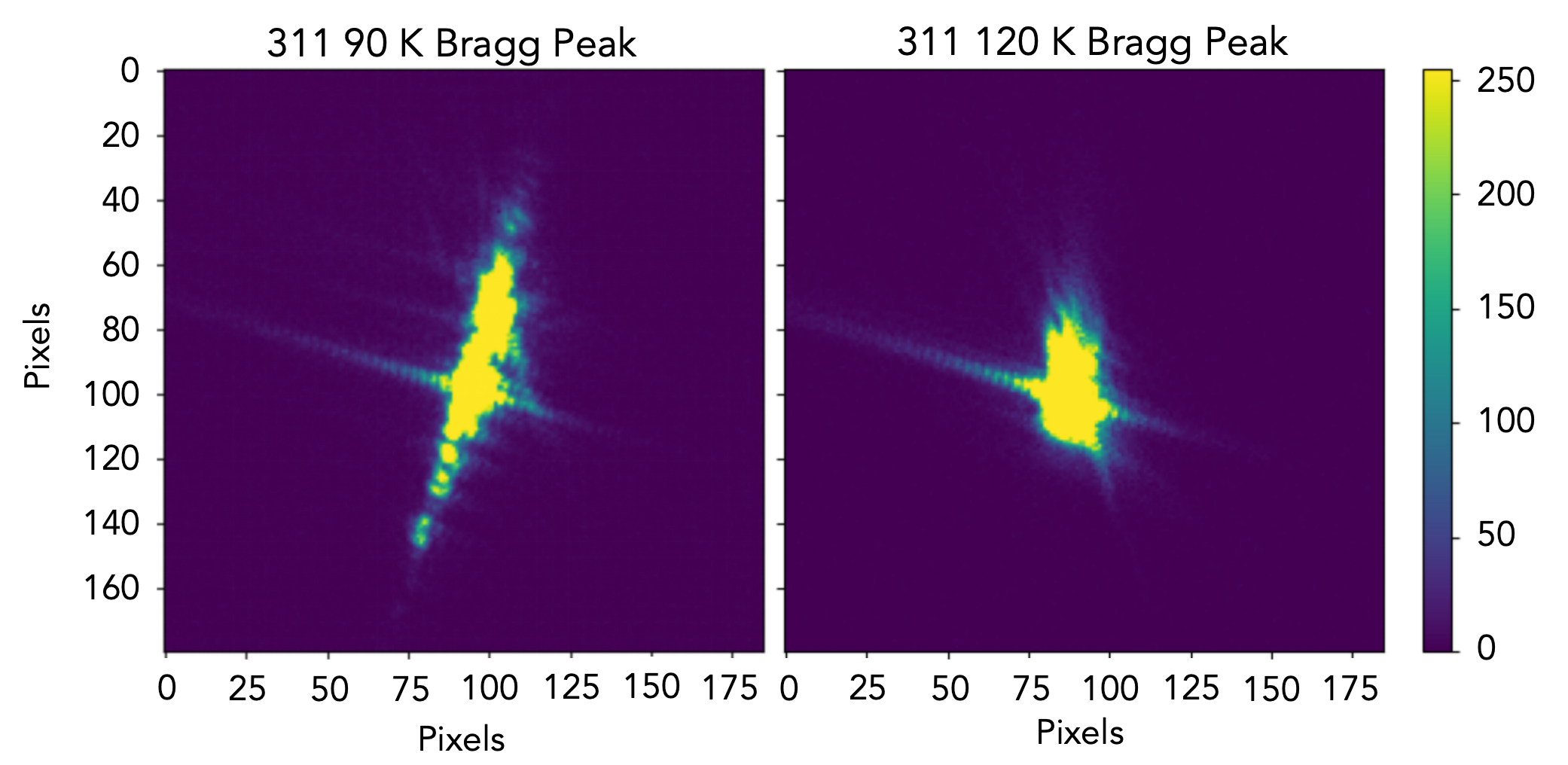} }}%
    \qquad
    \subfloat[\centering]{{\includegraphics[width=5.3cm]{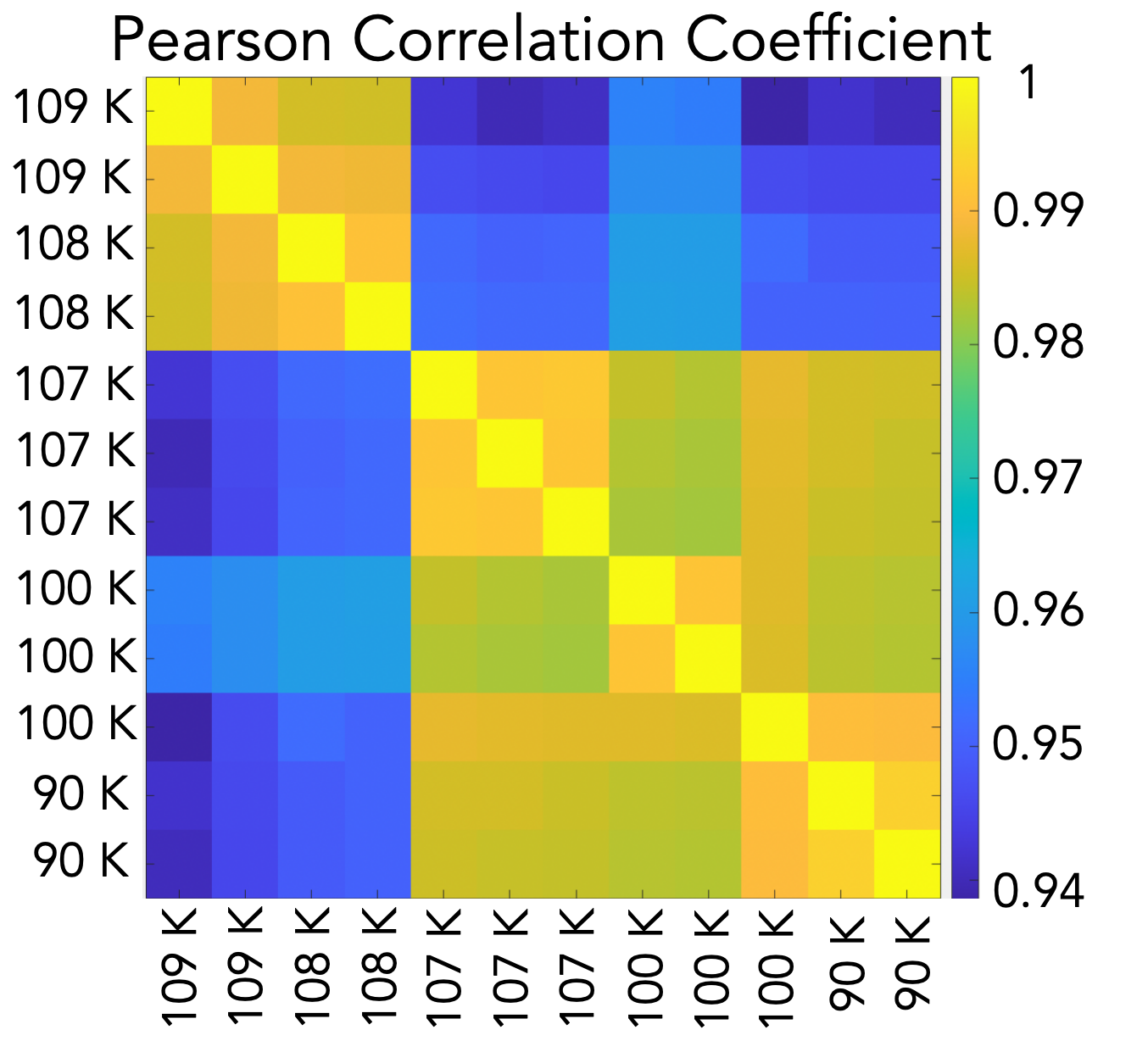} }}%
    \caption{The central slice of the 311 Bragg diffraction patterns at (a) 90 K and (b) 120 K, below and above the Verwey transition. (c) Pearson correlation coefficient between each of the 311 diffraction peaks after 3D subpixel alignment. }%
    \label{fig:2}%
\end{figure}

Figure $\ref{fig:2}$ shows the diffraction patterns of the 311 Bragg peak at 90 K and 120 K, below and above the Verwey transition. The splitting of the peak along the c-axis, roughly vertical, can be clearly seen at the lower temperature. The changes in the diffraction pattern with temperature can be further documented by calculating the Pearson Correlation Coefficient (PCC) between the measured 3D diffraction patterns after 3D subpixel alignment to the center of the array\cite{Guizar-Sicairos2008}. The PCC values between the different data for the 311 data at five temperatures, T = 109 K, 108 K, 107 K, 100 K, 90 K all below the Verwey transition, are shown as a “heat map” in Figure $\ref{fig:2}$(c). Each temperature set demonstrates high internal consistency, reflected by the high PCC values along the diagonal and between adjacent elements. The PCC heat map also significantly changes with temperature. There is a high similarity between the temperature sets at T = 107 K, T = 100 K, and T = 90 K. However, the cross-correlation is slightly lower for the temperature sets at T = 109 K and T = 108 K when compared to the aforementioned groups. This suggests minor variations between the T = 109 K, T = 108 K group and the T = 107 K, T = 100 K, and T = 90K group.

\begin{figure}%
    \centering
    \subfloat[\centering]{{\includegraphics[width=5.3cm]{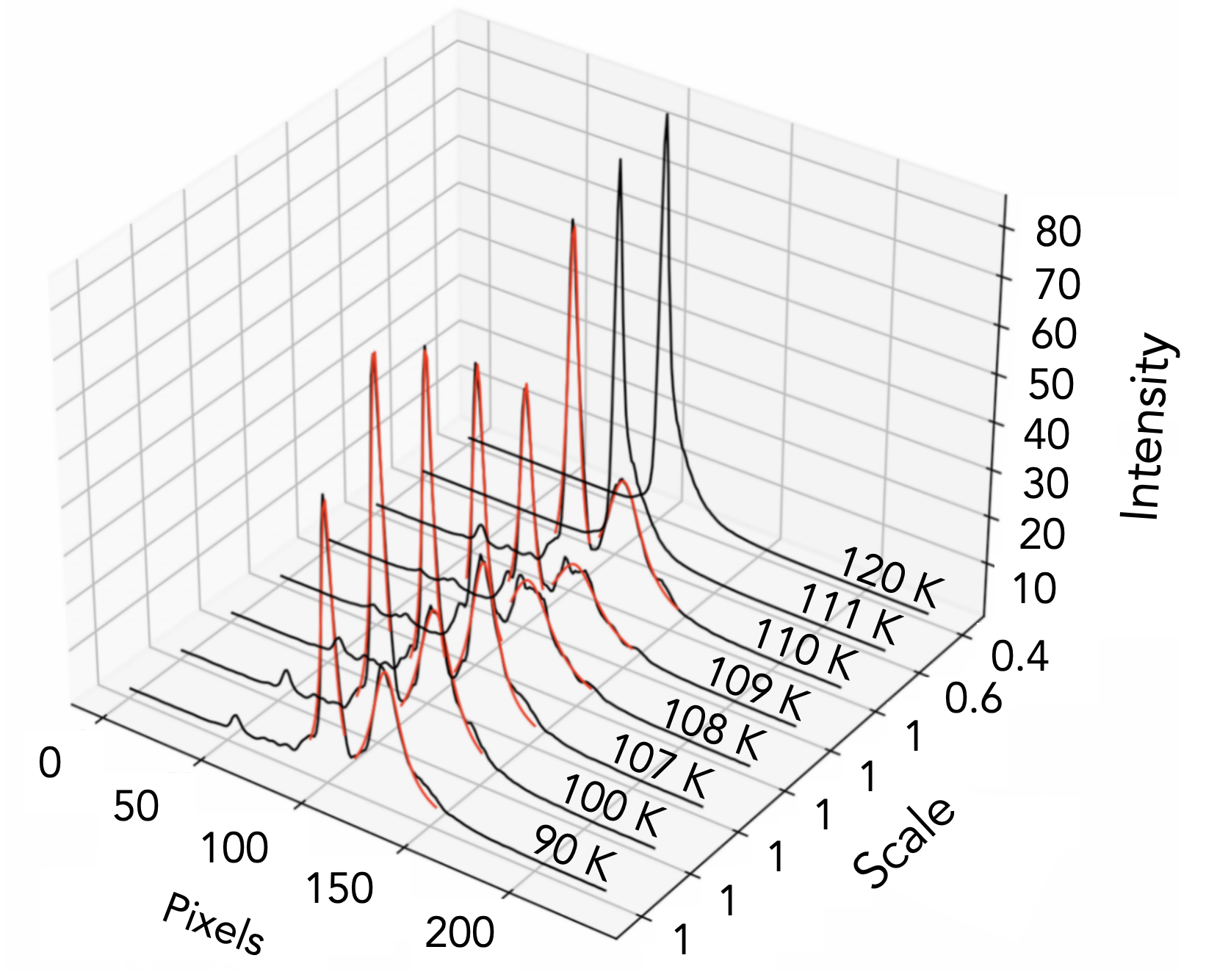} }}%
    \subfloat[\centering]{{\includegraphics[width=5.3cm]{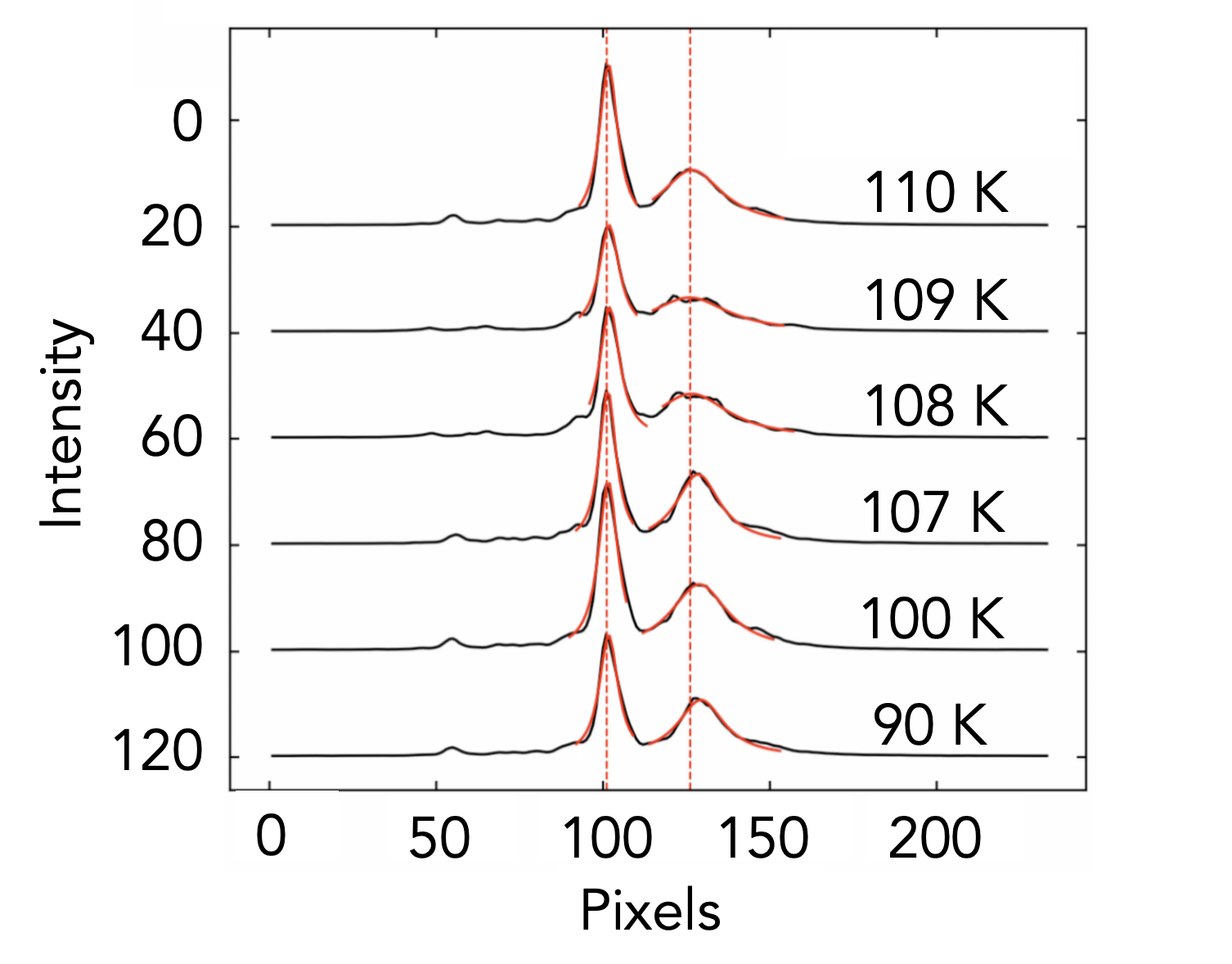} }}%
    \subfloat[\centering]{{\includegraphics[width=5.3cm]{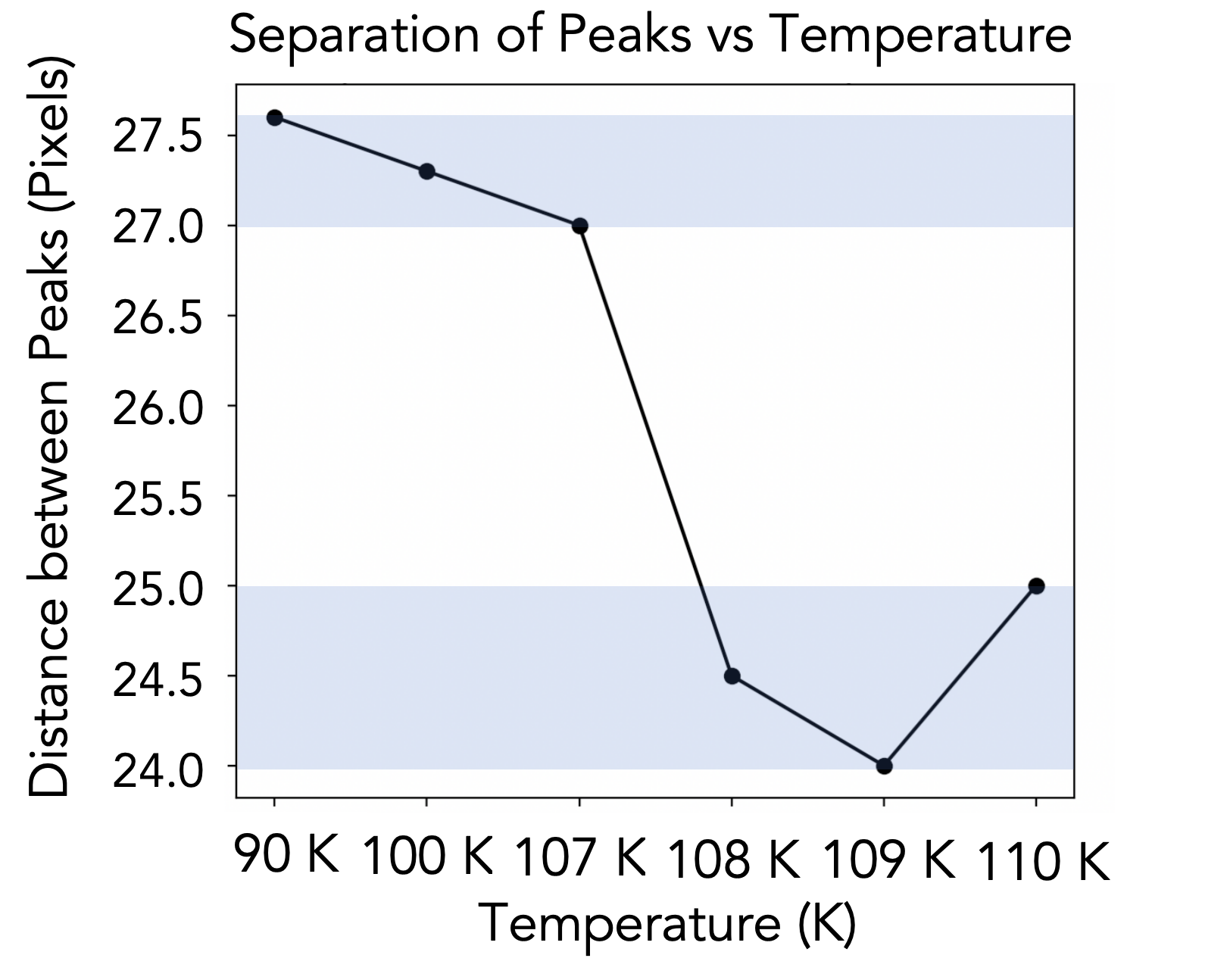} }}%
    \caption{Integrated 1D scan along the c-axis direction of the 311 reflection. (a) The integrated 1D scans are plotted in a 3D diagram, with the intensity scale labeled. Note that scans at T = 111 K and T = 120 K are scaled down respectively by a factor of 0.6 and 0.4. (b) The first maxima of each 1D scan are aligned vertically. The split peaks below the Verwey transition temperature are each fitted with Lorentzian curves. (c) Separation distance of the peaks on the detector seen in (b). }%
    \label{fig:3}%
\end{figure}

One-dimensional integrated scans of intensity along the c-axis of the 311 diffraction pattern are plotted in Figure $\ref{fig:3}$. A splitting of the bright peak was observed below 110 K, while the small temperature variations seen in the PCC between each scan of the two separated groups is visible as a change of distance between the split peaks, also shown in Table 1. From 90 K to 107 K, the distance between peaks remains relatively constant. From 108 K to 110 K, the distance between peaks decreases slightly. The angular separation of the two peaks is $\sim$0.07$^\circ$, roughly in agreement with the splitting of the 404 peak seen previously [16]. Above T$_{v}$, there is a single peak indicating an undistorted cubic crystal. 

We then reconstructed the diffraction patterns from the 311 reflections at different temperatures to visualize the domain structures associated with the peak splitting seen in the low-temperature monoclinic phase. In this work, the data were processed using the iterative phase retrieval methods in a genetic algorithm, known as “Guided Hybrid Input-Output” (GHIO)\cite{Chen2007}. For each reconstruction, the process involved 50 initial populations and 7 generations.  Each reconstruction begins with a randomly initialized guess, utilizing 50 distinct initial populations over 7 generations. In each generation, a sequence of 20 ER and 180 HIO iterations is performed with $\beta$ = 0.9, repeated three times, followed by an additional 20 ER iterations. The shrinkwrap algorithm\cite{Marchesini2003} was employed to periodically update the real-space support.

\begin{figure}
    \centering
    \includegraphics[width=0.9\linewidth]{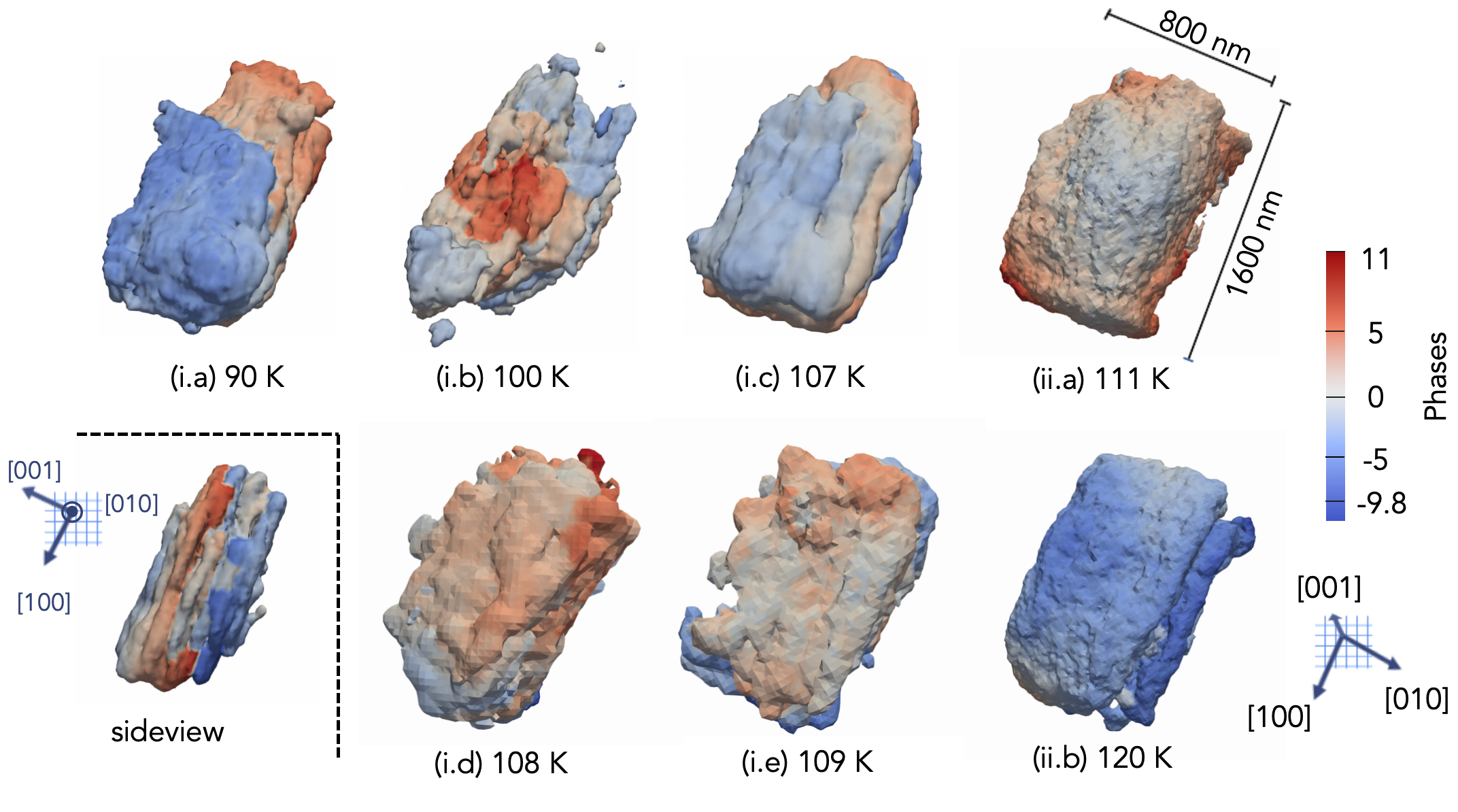}
    \caption{(a) Orientation of the sample mounted onto the sample stage with [100], [010], [001] directions labeled. (b) Reconstructions of the 311 reflection at T = 90 K, 100 K, 107 K, 108 K, 109 K, 111 K, and 120 K. Amplitude isosurfaces are colored by the local phase relative to the center of mass of the crystal, which is set to 0. All reconstructions are shown on the same scale. The Verwey transition was found to occur at 110 K for this sample. }
    \label{fig:4}
\end{figure}

The BCDI reconstructions in Figure $\ref{fig:4}$ show that the reconstructed crystal measures 1.6 $\mathrm{\mu m }$$ \times 0.8$  $\mathrm{\mu m }$ $\times$ 0.8 $\mathrm{\mu m }$, approximately one-fourth of the size of the original magnetite sample seen in Figure $\ref{fig:1}$. Post-experiment scanning electron microscopy (SEM) analysis confirmed that the sample remained superficially undamaged by X-ray beam exposure. This suggests there may be two (or more) crystals with different orientations adjacent to one another. No diffraction peaks from the other grain(s) of the FIB-cut block were encountered during any of the measurements, suggesting there is a misorientation between the measured crystal and the other grain(s). We conclude this was due to an accidental choice of the sample volume selected by FIB. 

The reconstructed crystal, mounted on its substrate, is shown in the laboratory coordinate system depicted in by the arrows in Figure $\ref{fig:4}$(b). In general, the external shape of the reconstructed sample, as indicated by the same normalized amplitude (threshold set to 0.05) of the reconstruction, appears block-like at various temperatures. However, above the Verwey transition, the reconstructed crystal exhibits smoother edges, whereas below the Verwey transition, a rougher, layered structure could be directly observed along the [001] direction along the crystal edges.

\begin{figure}[htbp]
    \centering
    \includegraphics[width=0.9\linewidth]{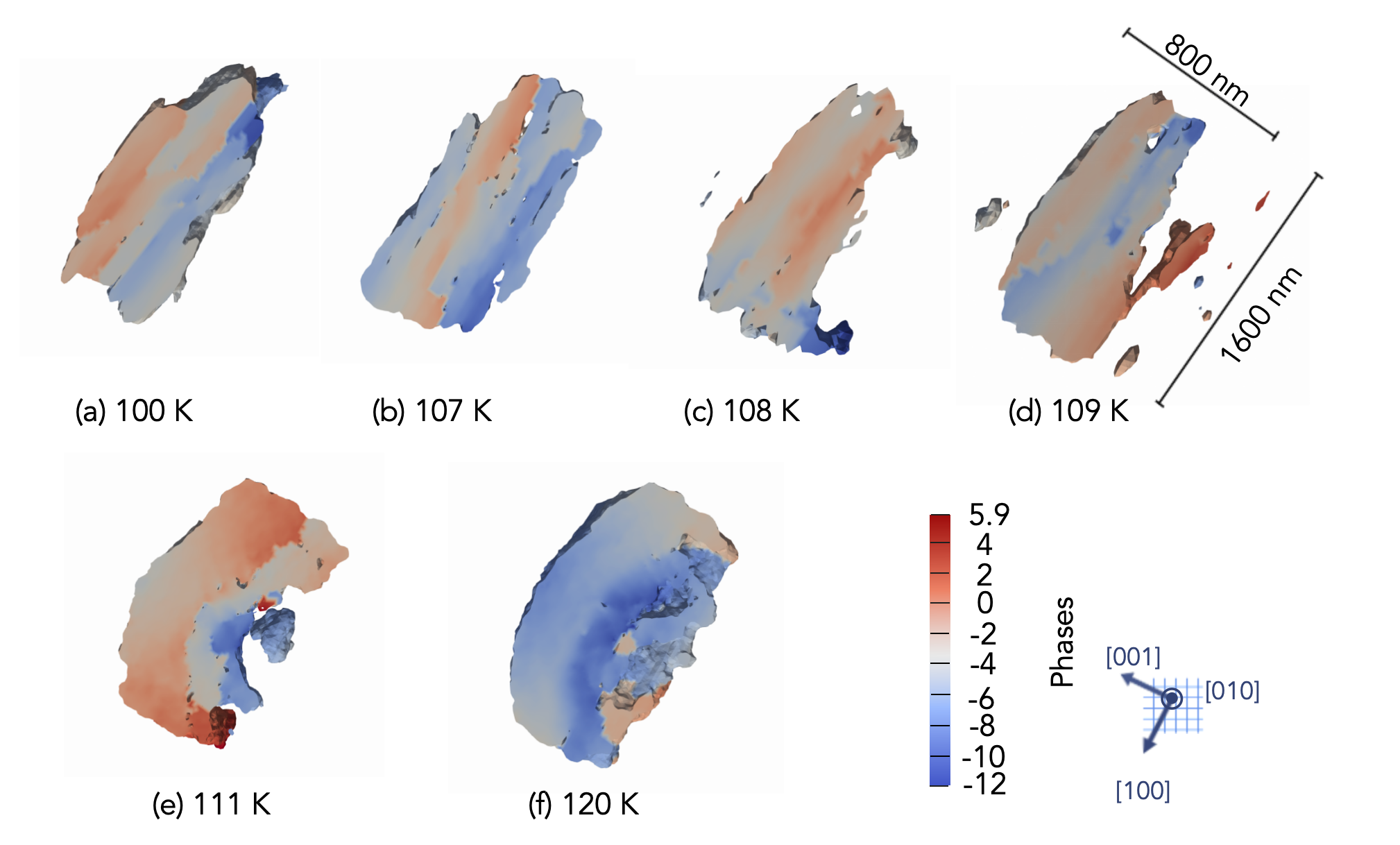}
    \caption{Slices of the BCDI phase in the XZ plane near the center of the magnetite sample. (a, b, c, d) are below the Verwey transition temperature. (e, f) are above the Verwey transition temperature. (e) and (f) depict a more uniform phase without stripes, where the relative phase remains consistent across the entire crystal. However, an offset in the absolute phase arises due to differences in the initial guesses used in the reconstruction.}
    \label{fig:5}
\end{figure}

As explained in the introduction, Eq($\ref{eq:1}$), the phases, encoded by color in Figure $\ref{fig:4}$, indicate the local displacements relative to the mean lattice parameter of the crystal. Here we interpret them as distortions due to the formation of domains within the underlying crystal. Phase unwrapping algorithms were applied due to the strong phase values observed within the magnetite sample in Figure $\ref{fig:4}$(b). The domains are seen more clearly in cross-sectional slices shown in Figure $\ref{fig:5}$. Clear stripe morphology domain structures are seen in the crystal phase image below the Verwey transition (Figures $\ref{fig:5}$(a-d)) while the phase is relatively flat above in Figures $\ref{fig:5}$(e, f). The images are shown as planes perpendicular to the substrate plane passing through the center of mass. The positive phase (red) represents the component of the displacement along the Q-vector, while the negative phase (blue) shows displacements opposite to the vector’s direction. The reconstructions of our magnetite sample exhibit strong striped phase characteristics, which required a phase unwrapping algorithm for resolving phase wraps extending above 2$\pi$\cite{Maier2015}. 

\begin{figure}[htbp]
    \centering
    \includegraphics[width=0.9\linewidth]{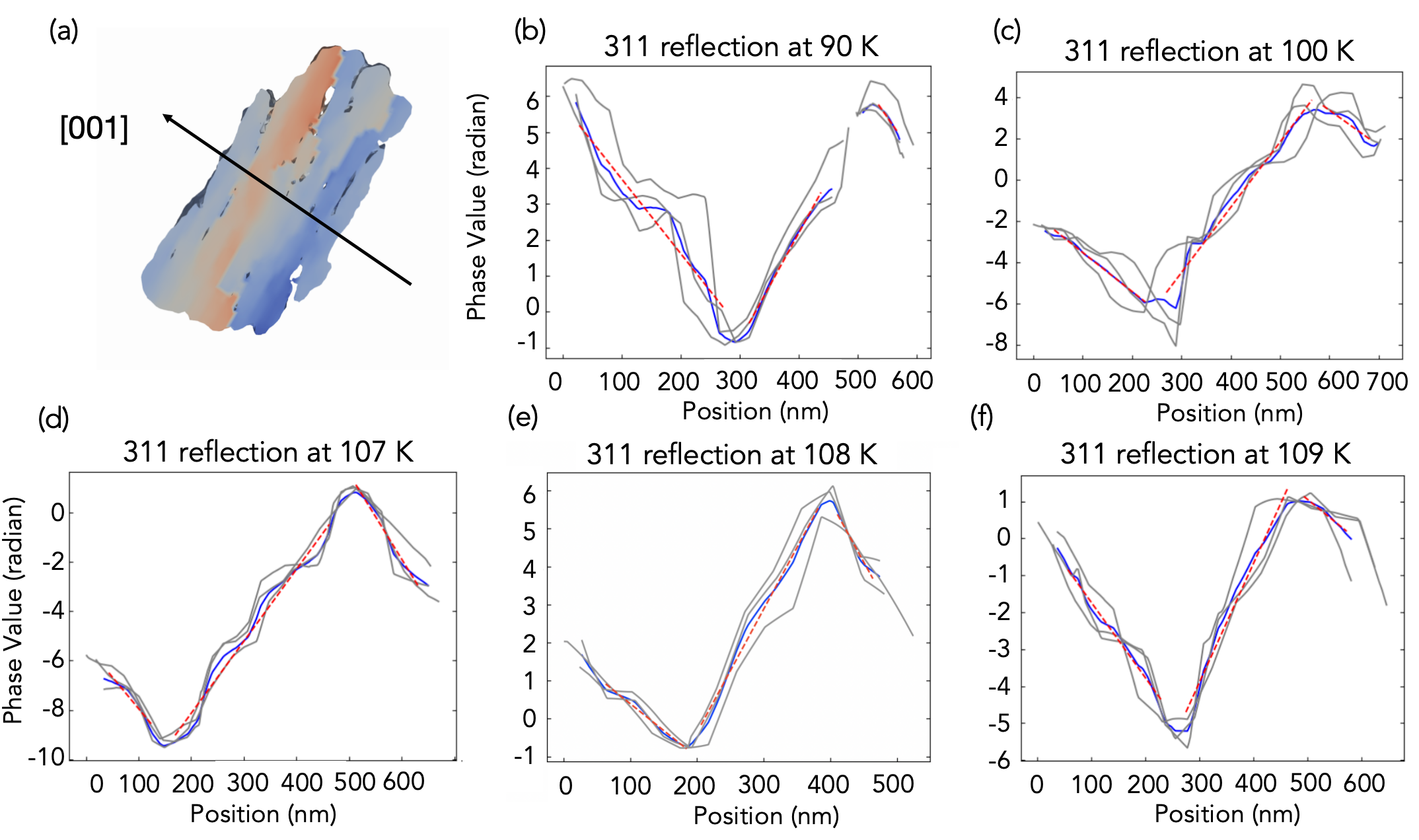}
    \caption{Phase profile along the [001] direction (indicated by the solid black line(a)) at (b) 90 K, (c) 100 K, (d) 107 K, (e) 108 K, and (f) 109 K. At each temperature, three curves are shown, representing line cuts at different positions in the crystal. Three gray line profiles are randomly chosen along the [001] direction, and the blue line profile is the average of the three gray line profiles. }
    \label{fig:6}
\end{figure}

Above the Verwey transition (111 K and 120 K), the phase remains relatively uniform across the entire crystal while changing abruptly into a layered “stripe” structure as the temperature decreases below the Verwey transition (110 K). In the two cross-sections observed above the Verwey transition, higher phase values are detected at the long ends of the crystal, with lower phase values near the surface and bottom. Below T$_{v}$, these layers are shifted in phase relative to one another along the [001] crystallographic direction, forming striped domains exhibiting a triangular wave pattern along the line profile plotted across the phase layers in Figure $\ref{fig:6}$. We fitted a linear triangular phase profile. The triangular cyclic pattern represents a special repetitive structure at low temperatures, uniquely aligned with the [001] axis, which has not been observed before. This choice of direction of the stripes has broken the cubic symmetry of the high-temperature phase.

The half-cycle of the triangular phase profile, representing the layer thickness, varies with temperature. Typically, three distinct layers span the magnetite sample in the BCDI image. The layer thickness exhibits notable variations with temperature shown in Table $\ref{tab:1}$, particularly around the Verwey transition temperature. At 90 K, the thickness is approximately 240 nm, increasing to 300 nm at 100 K, and reaching a peak of around 330 nm at 107 K. Interestingly, a sharp decrease in thickness is observed at higher temperatures, with values dropping to 210 nm at 108 K and further to 170 nm at 109 K. These oscillations in layer thickness might be indicative of the system's proximity to equilibrium within the magnetite crystal, which is influenced by the Verwey transition. The observed thickness oscillations likely arise from competing influences of thermal energy and the intrinsic structural reorganization associated with the Verwey transition, reflecting a delicate balance between the crystal's drive toward equilibrium and its response to external temperature changes.

\begin{table}[]
\centering
\begin{tabular}{cc}
\hline
Temperature (K) & Thickness of the layer (nm) \\ \hline
90              & 243.86 $\pm$ 10.1           \\
100             & 303.38 $\pm$ 33.4           \\
107             & 358.44 $\pm$ 10.6           \\
108             & 210.20 $\pm$ 2.31           \\
109             & 215.43 $\pm$ 42.5           \\ \hline
\end{tabular}
\caption{The measured thickness of the layer from the turning points(half cycle of the triangle phase profile) of the linear fits in Figure $\ref{fig:6}$. }
\label{tab:1}
\end{table}

A quantitative measure of the magnitude of crystal distortion from its high-temperature cubic state can be determined from phase gradients below the phase transition temperature. Linear fits of the phase profiles, overlaid in Figure $\ref{fig:6}$, were used to measure the phase gradient along the [001] direction and listed in Table $\ref{tab:2}$. The gradient values with alternating signs reflect the opposite directions of the triangle wave. Furthermore, the asymmetrical nature of the triangular wave fitting of the phase profile reveals a non-uniform phase variation along the [001] direction, corresponding to the layered stripes' orientation. 

\begin{figure}
    \centering
    \includegraphics[width=0.5\linewidth]{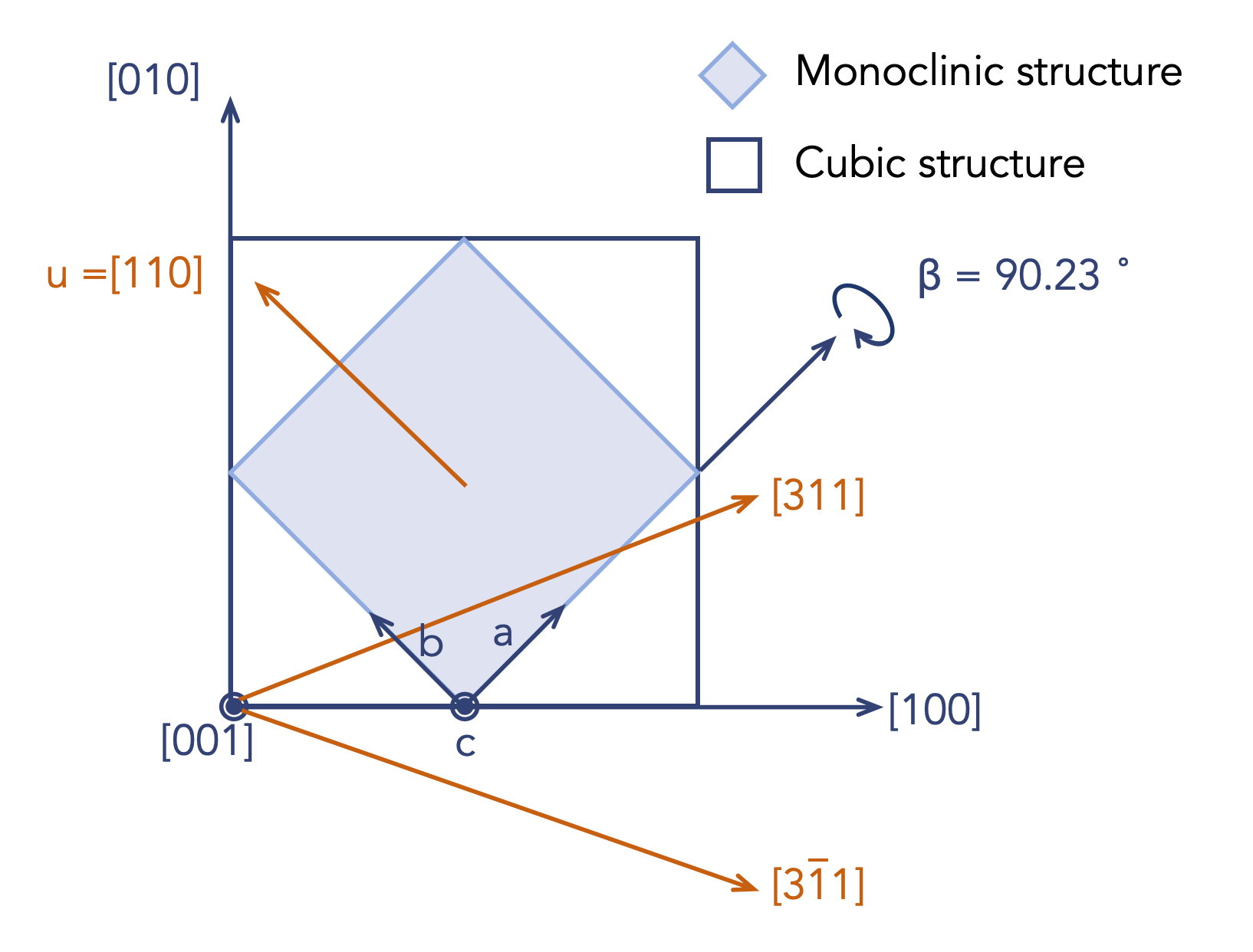}
    \caption{Definition of the monoclinic unit cell of magnetite at low temperature in the pseudocubic coordinate system used in this paper. The distortion corresponds to a rotation around the b axis indicated. The resulting crystal displacement, u,  is projected onto the Q-vector of the two possible 311 reflections, inclined to the plane of the page, as shown.}
    \label{fig:7}
\end{figure}

To relate these measured slopes to the structural distortion of the crystal expected at low temperatures, we refer to Figure $\ref{fig:7}$. The low-temperature monoclinic unit cell was determined to be a = 5.944 $\mathring{A}$, b = 5.925 $\mathring{A}$, c = 16.775 $\mathring{A}$, and $\beta$ = 90.23$^\circ$\cite{Wright2002}. The in-plane axis lengths are $\sqrt{2}$ shorter than the cubic reference and rotated 45$^\circ$, as shown. There are two permutations of the a and b axis directions and two possible directions of the $\beta$ angle. In Figure $\ref{fig:7}$, we account for all these possibilities with a single b-axis direction and two possible {311} measurement reflections. The monoclinic $\beta$ angle distorts the unit cell by introducing a shear along the [-110] direction marked in Figure $\ref{fig:7}$. At a given height z in the structure the magnitude of the shift due to shear is u = z / tan$\beta$, with a component along the two possible {311} reciprocal space vectors Q = 2$\pi$/a$_{0}$ $\cdot$ (3,$\pm$1,1), where a$_{0}$ is the cubic lattice constant, 8.397 $\mathring{A}$. According to Eq($\ref{eq:1}$), the phase shift is therefore: 

\begin{equation}
    \Delta \phi = \vv{u}(\vv{r}) \cdot \vv{Q} = \frac{2\pi z(3\pm1)\mathrm{cos}(\beta)}{\sqrt{2}a_{0}\mathrm{sin}(\beta)}
\end{equation}

and the phase slope with respect to height in the structure is given by:

\begin{equation}
    \frac{\mathrm{d}\phi}{\mathrm{d}z}= \frac{2\pi(3\pm1)\mathrm{cos}(\beta)}{\sqrt{2}a_{0}\mathrm{sin}(\beta)}
\end{equation}

This equation allows us to obtain the effective monoclinic angle of the local distortions from the value of the phase slope in Table $\ref{tab:2}$. 

\begin{equation}
    \beta = 2\pi - \mathrm{arctan}(\frac{2\pi(3\pm1)}{\sqrt{2}a_{0} \frac{\mathrm{d}\phi}{\mathrm{d}z}})
\end{equation}

For a given $\beta$ angle at the 311 reflection, we would expect there to be spatially separated domains with two possible phase slopes, due to the different coupling of the b-axis with two possible {311} directions (or vice versa). In our observations, however, only one type of domain was detected, extending over the entire sample. Without additional information, we assumed the “-” sign in Table $\ref{tab:2}$ to get the larger of the two possible $\beta$ angle values, which was still slightly smaller than the 90.23$^\circ$ value previously determined by Rietveld refinement\cite{Wright2002}. Table $\ref{tab:2}$ also shows there was no significant temperature variation.

\begin{table}[]
\centering
\begin{tabular}{cccc}
\hline
Temperature (K) & Gradient (rad/nm)                                                  & average (Gradient) & Monoclinic Angle $\beta$ ($^\circ$) \\ \hline
90              & \begin{tabular}[c]{@{}c@{}}-0.0211\\ 0.0305\\ -0.0203\end{tabular} & 0.0256             & 90.139 $\pm$ 0.018                  \\
100             & \begin{tabular}[c]{@{}c@{}}-0.0196\\ 0.0324\\ -0.0209\end{tabular} & 0.0262             & 90.142 $\pm$ 0.022                  \\
107             & \begin{tabular}[c]{@{}c@{}}-0.0365\\ 0.0294\\ -0.0352\end{tabular} & 0.0326             & 90.177 $\pm$ 0.012                  \\
108             & \begin{tabular}[c]{@{}c@{}}-0.0141\\ 0.0325\\ -0.0318\end{tabular} & 0.0277             & 90.153 $\pm$ 0.012                  \\
109             & \begin{tabular}[c]{@{}c@{}}-0.0217\\ 0.0309\\ -0.0133\end{tabular} & 0.0242             & 90/131 $\pm$ 0.028                  \\ \hline
\end{tabular}
\caption{The measured phase gradients of the linear fits in Figure $\ref{fig:6}$, and the derived monoclinic distortion angle. In each case, the quoted “average” value is the average of the gradient magnitudes. This accounts for the possible phase ramps appearing in the reconstructed images.}
\label{tab:2}
\end{table}

\section{Discussion}
In our experiment, the phase transition was found to occur at 110 K, whereas the Verwey transition is reported to be at T$_{v}$ = 125 K\cite{Kirschvink1992}\cite{Kobayashi2024}\cite{Fatimah2021}\cite{Verwey1939}\cite{Verwey1936}\cite{Joaquin2004}\cite{Friedrich2002}. The measurement of electric resistivity vs temperature and in situ TEM cooling experiment on our sample both showed that the Verwey transition temperature is 115 K. The temperatures quoted throughout the paper are those measured during the experiment, but, as noted above, there may be a small difference due to positioning of the thermometer. However, it is also reported that the Verwey transition temperature can vary with external factors such as oxidation. Research indicates that the slow oxidation of magnetite nanoparticles can shift this transition temperature\cite{Kim2021}. Additionally, it was explored how environmental and natural variations among magnetite minerals contribute to a distribution in Verwey transition temperatures\cite{Mike2021}. The Verwey transition in natural and synthetic magnetite may be influenced by both chemical and structural properties, including nanoscale material formulation\cite{Meyers2006}.

\begin{figure}%
    \centering
    \subfloat[\centering]{{\includegraphics[width=7.8cm]{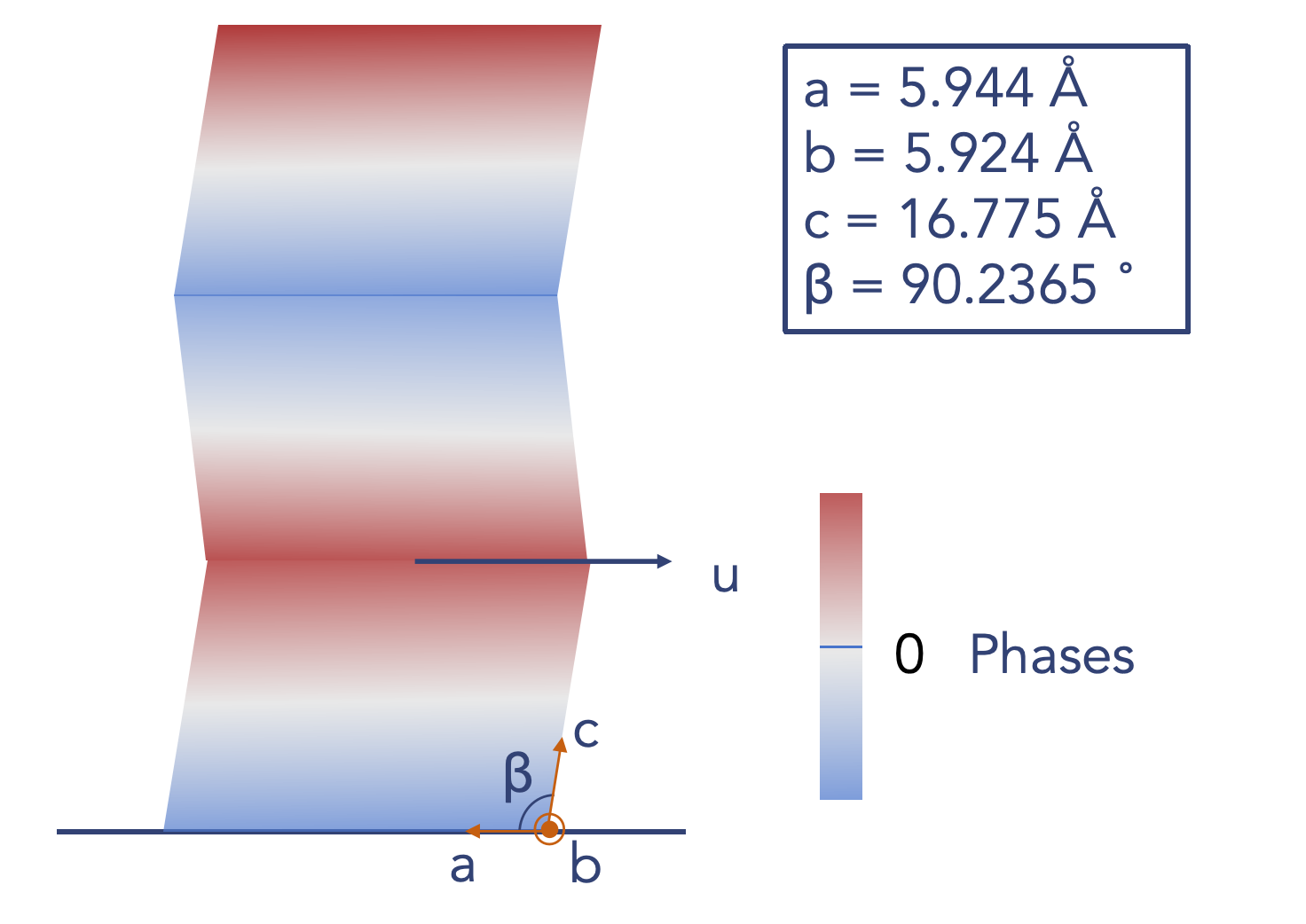} }}%
    \qquad
    \subfloat[\centering]{{\includegraphics[width=7.8cm]{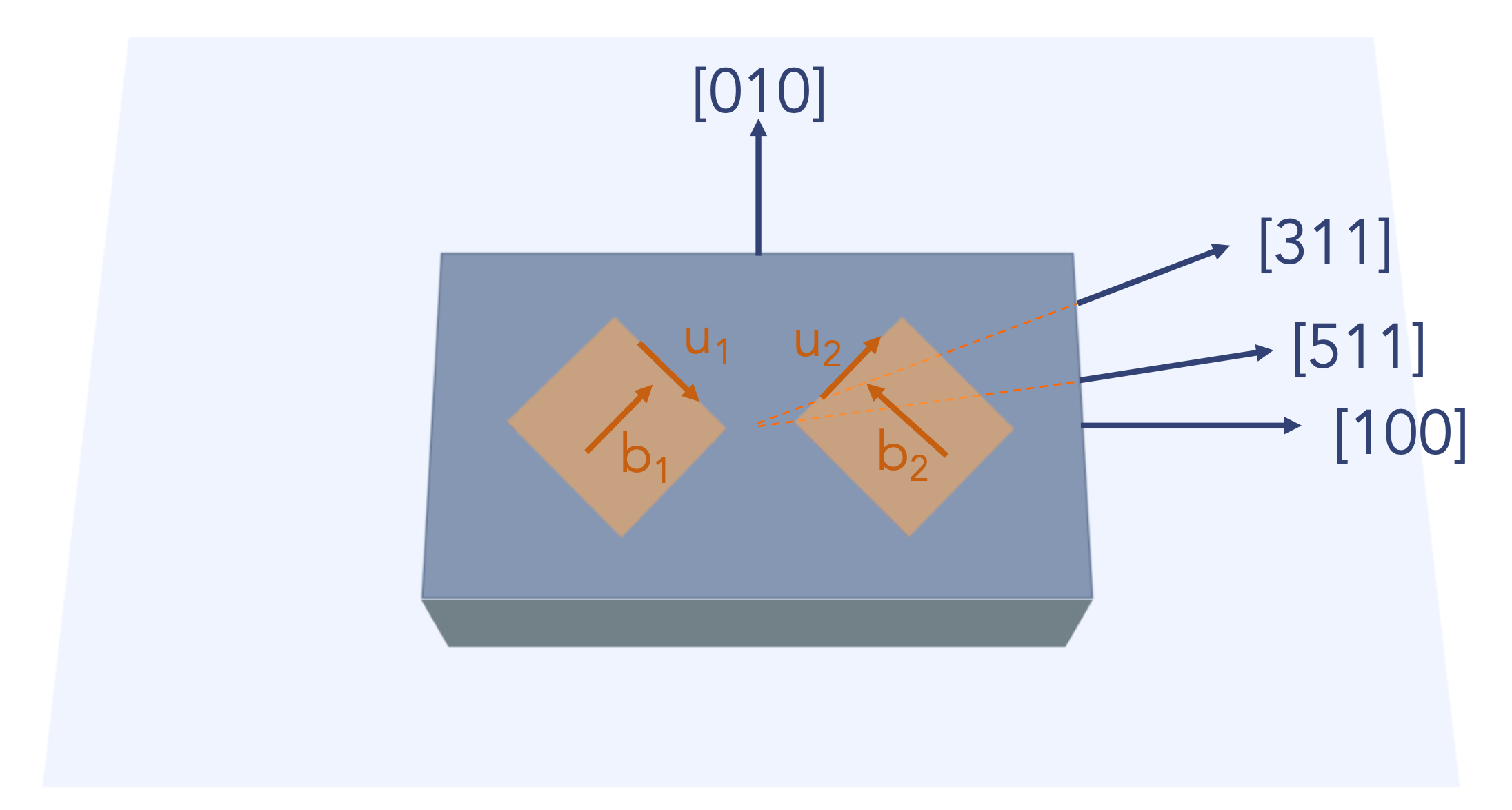} }}%
    \caption{(a) The monoclinic unit cell structure is layered along [001] direction with an alternating direction between the twins. (b) two possible directions that the monoclinic a and b axes can line up with the cubic directions of the high-temperature phase.}%
    \label{fig:8}%
\end{figure}

A one-dimensional layered structure of phase domains of the low-temperature monoclinic phase of Fe$_{3}$O$_{4}$ was imaged parallel to the [001] upper surface plane of the FIB-mounted crystal. These stripes follow the classical Martensitic twin morphology\cite{Shuvalov1988} shown in Figure $\ref{fig:8}$(a) where the monoclinic angle alternates direction between the twins. The formation of a striped morphology minimizes the overall shape change of the crystal because alternating stripes return the crystal to its undistorted configuration. The BCDI phase measurement records the displacement from the average crystal structure, which is seen to form a V-shaped profile as a function of depth in the crystal. The slope of the phase vs depth is a measure of the monoclinic angle seen in our sample. The monoclinic angle $\beta$ was found to vary from 90.13$^\circ$ to 90.18$^\circ$ $\pm$ $\sim$0.02$^\circ$ with temperatures. While this variation is at the level of the experimental error, the average value is significantly smaller than the published monoclinic angle, 90.23$^\circ$\cite{Wright2002}.

The observed spacing of twins in the Martensitic domain structure, measuring approximately 240 nm at 90 K and around 210 nm at 107 K, reflects a delicate balance between the excess energy of twin boundaries and the elastic energy required to distort the crystal as a whole. Above the Verwey transition, the martensitic domain structure collapses into one domain, driven by the thermal activation of domain wall movement. The variations in spacing, particularly associated with the phase transition, result from a compromise and balancing of the thermal energy due to external temperature change and intrinsic structural reorganization within the magnetite crystal within the twin boundary. 

What is notable about the domain structure seen in this experiment is that only one stripe structure is formed along the surface normal [001] axis. No stripes were seen that were associated with the [100] or [010] cubic axis directions. This apparent symmetry breaking might arises because strain relief is easier in the surface-normal direction which allows free displacement of the open crystal edges to accommodate the lateral strains.

\begin{figure}
    \centering
    \includegraphics[width=0.8\linewidth]{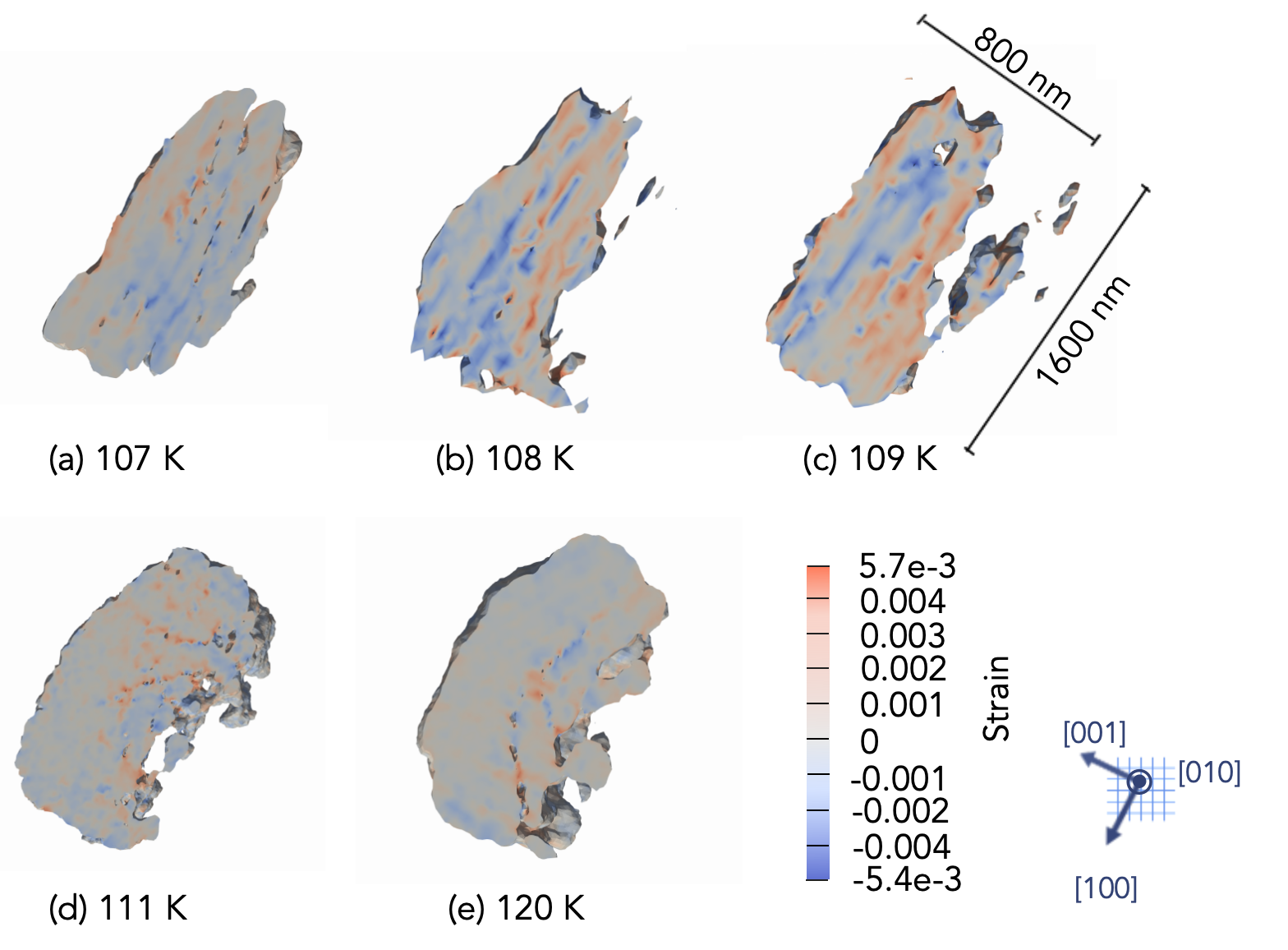}
    \caption{A central slice through the strain distribution at (a)107 K, (b)108 K, (c)109 K, (d)111 K and (e)120 K. }
    \label{fig:9}
\end{figure}

The strain distributions exhibit clear boundaries that are parallel to the layered phase distribution seen in the reconstructed crystal images shown in Figure $\ref{fig:9}$. The nano-strain is defined as the derivative of the image phase with respect to distance along the direction of the Q-vector [38]. Figure $\ref{fig:9}$ shows a comparison of strain distribution found along [311] direction at 107 K, 108 K, 109 K, 111 K and 120 K. At higher temperatures (120 K and 111 K), the strain distribution appears more uniform, suggesting a smoother internal stress state. At lower temperatures below the Verwey transition, the strain distribution becomes increasingly fragmented, indicating localized stress concentrations due to internal layered phase variations. This evolution highlights the temperature-dependent mechanical behavior of the crystal, where higher temperatures allow for stress relaxation, while lower temperatures promote strain localization. We suspect that any stripes that could have formed on the sides of the crystal would be constrained by the bottom side of the sample being attached to the Pt weld holding the FIB-cut crystal to the substrate. Thus, we think that the symmetry breaking is associated with anisotropies introduced by the mounting of the crystal.

A more subtle breaking of the symmetry is that only one monoclinic domain appears to form on the [001] surface of the crystal. As mentioned above, only one set of stripes was detected along the c-axis direction, corresponding to the distortion direction and the 3-11 reflection shown in Figure $\ref{fig:7}$. As shown in Figure $\ref{fig:8}$(b), there are two equivalent directions that the monoclinic a and b axes can line up with the cubic directions of the high-temperature phase, yet only one of them is chosen. This might be related to the rectangular shape of the crystal or somehow due to the contact with other grain(s) in the sample that was seen in SEM but not imaged by BCDI, but this is not obvious from the 45$^\circ$ rotation of the axes. It is likely there is some residual strain in the crystal left by the FIB sample preparation, which is responsible for pinning the choice of monoclinic domain. We observed the same stripe direction after cycling the temperature several times through T$_{v}$, so this symmetry breaking is not a random choice. Our findings provide deeper insights into the fundamental behavior of magnetite and similar materials undergoing structural transitions, emphasizing the intricate relationships between energy minimization, thermal effects, and crystal stability.

\section*{Acknowledgments}
Work at Brookhaven National Laboratory was supported by the U.S. Department of Energy, Office of Science, Office of Basic Energy Sciences, under Contract No. DE-SC0012704. Work performed at UCL was supported by EPSRC. Beamtime at I16 was provided by Diamond Light Source under proposal MM36376-1. The FIB sample preparation was supported by the resources of Center for Function Nanomaterials (CFN) at BNL. The authors would like to acknowledge the use of the University of Oxford Advanced Research Computing (ARC) facility in carrying out this work\cite{Richards2015}.

%Bibliography
\bibliographystyle{unsrt}  
\bibliography{references}  
\cite{Gleitzer1985} \cite{Liu2015} \cite{Lee2008} \cite{Senn2011} \cite{Lorenzo2008} \cite{Kirschvink1992} \cite{Kobayashi2024} \cite{Fatimah2021} \cite{Verwey1939} \cite{Verwey1936} \cite{Joaquin2004} \cite{Friedrich2002} \cite{Bragg1915} \cite{Lizumi1982} \cite{Wright2011} \cite{Wright2002} \cite{Shuvalov1988} \cite{Perversi2019} \cite{Meyers2006} \cite{Tabis2009} \cite{Lindquist2019} \cite{Wright2000} \cite{Robinson2009} \cite{Hofmann2017} \cite{Pfeifer2006} \cite{Fienup1978} \cite{Chen2007} \cite{Yu2024} \cite{Wu2021} \cite{Masto2024} \cite{Yang2021} \cite{Clark2015} \cite{Liu2022} \cite{Atlan2023} \cite{Diao2020} \cite{Yuan2019} \cite{Mokhtar2024} \cite{Dhariwal2024} \cite{Hofmann2023} \cite{Busing1967} \cite{Sayre1952}  \cite{Guizar-Sicairos2008} \cite{Marchesini2003}
\cite{Maier2015} \cite{Kim2021} \cite{Mike2021}  \cite{Richards2015}

\end{document}